# Evidence of Pure Spin-Current Generated by Spin Pumping in Interface Localized States in Hybrid Metal-Silicon-Metal Vertical Structures

C. Cerqueira[1,6*], J. Y. Qin[2,5*], H. Dang[1], A. Djeffal[2], J.-C. Le Breton[3], M. Hehn[2], J.-C. Rojas-Sanchez[2], X. Devaux[2], S. Suire[2], S. Migot[2], P. Schieffer[3], J.-G. Mussot[2], P. Laczkowski[1], A. Anane[1], S. Petit-Watelot[2], M. Stoffel[2], S. Mangin[2], Z. Liu[4], B. W. Cheng[4], X. F. Han[5], H. Jaffrès[1a)], J.-M. George[1b)], Y. Lu[2c)]

[1]Unité Mixte de Physique, CNRS, Thales, Univ. Paris-Sud, Université Paris-Saclay, 91767, Palaiseau, France
[2]Institut Jean Lamour, UMR 7198, CNRS-Université de Lorraine, Campus ARTEM, 2 Allée André Guinier, BP 50840, 54011 Nancy, France
[3]Univ Rennes, CNRS, IPR (Institut de Physique de Rennes) - UMR 6251, F-35000 Rennes, France
[4]State Key Laboratory on Integrated Optoelectronics, Institute of Semiconductors, University of Chinese Academy of Sciences, Chinese Academy of Sciences, Beijing 100083, P. R. China
[5]Beijing National Laboratory of Condensed Matter Physics, Institute of Physics, University of Chinese Academy of Sciences, Chinese Academy of Sciences, Beijing 100190, P. R. China
[6]Laboratoire des Solides Irradiés, École Polytechnique, CNRS, CEA, Université Paris-Saclay, 91128 Palaiseau, France

* Authors have equivalent contribution.
Corresponding authors:
a) *henri.jaffres@cnrs-thales.fr*
b) *jeanmarie.george@cnrs-thales.fr*
c) *yuan.lu@univ-lorraine.fr*

**Abstract**

Due to the difficulty to grow high quality semiconductors on ferromagnetic metals, the study of spin diffusion transport in Si was only limited to lateral geometry devices. In this work, by using ultra-high vacuum wafer-bonding technique, we have successfully fabricated metal-semiconductor-metal CoFeB/MgO/Si/Pt vertical structures. We hereby demonstrate pure spin-current injection and transport in the perpendicular current flow geometry over a distance larger than 2μm in *n*-type Si at room temperature. In those experiments, a pure propagating spin-current is generated via ferromagnetic resonance spin-pumping and converted into a measurable voltage by using the inverse spin-Hall effect occurring in the top Pt layer. A systematic study by varying both Si and MgO thicknesses reveals the important role played by the localized states at the MgO/Si interface for the spin-current generation. Proximity effects involving indirect exchange interactions between the ferromagnet and the MgO/Si interface states appears to be a prerequisite to establish the necessary out-of-equilibrium spin-population in Si under the spin-pumping action.

**Introduction**

During the last few years, silicon-based semiconductor spintronics[1] has gained a growing interest owing to the potential to overcome Moore's law and its potential integration into the mainstream of Si technology. Not only Si is one of the most common elements in the earth, but also it is a perfect candidate for spin transport due to the possibility to maintain its spin memory over micrometric distances[1]. This is due to several features: *i*) the inversion symmetry of its crystalline structure minimizing D'yakonov-Perel (DP) interactions and related spin-relaxation[2]; *ii*) its small atomic number minimizing the spin-orbit coupling[3] and subsequent spin-depolarization; *iii*) as well as the very high weight (95%) of $^{28}$Si isotope with zero nuclear spin and the reduced hyperfine interactions[4]. The development of silicon spintronics is guided by the ability to generate and manipulate the spin information before its electrical conversion, whose operations constitute the building block for most of the envisioned spin-based semiconductor devices. During the last decade, several proof-of-concepts have been achieved including spin-current injection and conversion in Si using ferromagnet/tunnel barrier[5,6,7], thermal gradient[8], spin pumping[9] or spin-to-charge conversion by inverse spin Hall effect (ISHE)[10]. Until now, investigations of pure spin-current (associated to zero charge current) were mainly focused on the local and non-local spin-valve effects in lateral devices, as demonstrated in a couple of recent papers[6,11,12]. The main reason is the difficulty to grow high quality semiconductor (SC) on ferromagnetic transition metals (FM). The chemical wetting problem, the nature of bonds and their strong lattice mismatch between SC and FM have prohibited the design of vertical metal-semiconductor-metal structures. However, the lateral devices show many drawbacks, such as complicated lithography fabrication processes and geometry dependent spin transport. The spin current has to be firstly injected perpendicularly from the injector part, before propagating laterally inside the channel, and this may introduce some spurious effects on the results. For instance, interfacial spin relaxation[13] in the lateral channel may alter the bulk semiconductor properties leading to an underestimation of the spin diffusion length (SDL). Although electron spin injection and conversion were already demonstrated with hot electron transport in vertical structures[14,15,16,17], the demonstration of a large diffusive pure spin-current is still lacking.

So far, spin pumping (SP) is proven to be an efficient method[18,19] to create a pure spin-current for spin-injection from FM to SC. SP is generated by radio frequency (RF) excitation of a ferromagnet in magnetic resonance (FMR). Combined with a heavy metal (like Pt, W or Pd) for the spin-to-charge conversion by ISHE, the spin-current can be electrically probed in metals[19,20,21,22], semiconductors[9,23], organic materials[24], or more recently in topological insulators[25,26]. By employing SP and ISHE, a clear spin transport in *p*-Si[9] and Ge[23] has been reported in the lateral devices. In the present work, by combining SP-injection and ISHE (SP-ISHE) measurements, we demonstrate the establishment of a pure spin-current over micrometric distance in *n*-Si at room temperature in CoFeB/MgO/Si/Pt hybrid vertical structures fabricated by wafer-bonding technique. By varying the Si thickness, we were able to estimate the spin-diffusion length in *n*-Si. Moreover, our results clearly evidence that the MgO/Si interface localized states and the indirect exchange proximity play a major role in the spin-pumping process as a prerequisite for spin-current generation and propagation through the Si interlayer.

**Results and discussions**

Figure 1a schematically shows the fabricated layer stack consisting of *n*-Si//MgO ($t_{MgO}$)/CoFeB (5.2)/MgO ($t_{MgO}$)/*n*-Si ($t_{Si}$)/Pt(6) (with thickness in nanometer). Please see Methods and Supplementary Note 1 for more details. In order to check the quality of the wafer-bonding, high-resolution scanning transmission electron microscopy (HR-STEM) is employed to observe the interfacial structure in the bonding region. Figure 1b reveals very flat and sharp interfaces between CoFeB, MgO and Si. One does not observe any pinholes in the middle of the CoFeB bonding region (pointed at with red arrow). Although CoFeB and MgO layers appear to be amorphous, the two Si layers are perfectly monocrystalline, as indicated by the FFT diffraction patterns (see insets of Figure 1b). Moreover, the two Si layers can be controlled to have exactly the same crystalline orientations. However, one cannot exclude the presence of localized defects such as oxygen vacancies in the MgO barrier[27].

**Demonstration of spin-current injection by spin pumping**

SP-ISHE experiments were carried out to demonstrate the spin-injection efficiency. FMR is firstly performed on the layer stack containing a 5.2nm thick CoFeB layer. The inset of Figure 2a displays the typical FMR spectra acquired at 8 GHz for the CoFeB (5.2nm)/MgO (2.2nm)/Si (3μm)/Pt (6nm) sample.

By using a standard fitting procedure (see Supplementary Note 2), we were able to extract the damping parameter $\alpha$ of the 3μm thick Si sample which is about $5.1\times10^{-3}$. This value is much lower than that of the CoFeB/Pt sample ($9.6\times10^{-3}$), but larger than that of a CoFeB/Al ($1.83\times10^{-3}$) reference sample grown on the SiO$_2$ substrate, which is normally free of any spin-current dissipation. The latter constitutes an upper bound for the intrinsic damping $\alpha_0$ of a 5.2nm thick CoFeB layer. This generally reveals a certain spin-current dissipation in the Si sample scaling with the difference of the damping $\Delta\alpha=\alpha-\alpha_0$ and expressed via the spin-mixing conductance $g_{sf}^{\uparrow\downarrow} = \frac{4\pi\Delta\alpha M_{eff} t_{CoFeB}}{g_L \mu_B}$ where $g_L$, $\mu_B$, $M_{eff}$, and $t_{CoFeB}$ are, respectively, Landé factor, Bohr magneton, effective magnetization and CoFeB layer thickness.

In order to evaluate the spin transport through Si, the ISHE voltage on the top Pt layer is acquired. Figure 2a displays a typical spectrum of the electromotive force (EMF) acquired at the same resonance frequency (8GHz) for the layer stack containing a 3μm thick Si layer. The EMF spectrum is composed of two parts[18,20], a symmetric Lorentzian shaped signal plus an asymmetric signal, which can be decomposed into $V(H) = V_{\text{offset}} + V_{\text{sym}} \frac{\Delta H^2}{(H-H_{\text{res}})^2+\Delta H^2} - V_{\text{asym}} \frac{\Delta H(H-H_{\text{res}})}{(H-H_{\text{res}})^2+\Delta H^2}$. $V_{\text{sym}}$ and $V_{\text{asym}}$ are the magnitude of the symmetric and antisymmetric voltage contributions, respectively. $H_{\text{res}}$ is the resonant field, and $\Delta H$ is the width of the EMF peak. Compared to the study in lateral p-Si device[9], the EMF signal in the vertical device is characterized by a rather large antisymmetric contribution. The different ratio between the symmetric and antisymmetric components could be related to the different relative phase between RF electric and magnetic field, which strongly depends on the material losses as well as the coplanar wave guide (CPW) characteristics[28,29].

In order to exclude a possible role played by lateral thermal gradients[19,30], the EMF signal as a function of RF-field power is acquired on one sample with a 37nm thick Si interlayer, as shown in Figure 2b. Obviously, the intensity of the EMF signal increases with increasing RF-power, $P_{\text{RF}}$. After extraction of the $V_{\text{sym}}$ and the $V_{\text{asym}}$ components, we obtain a very good linear relationship between ($V_{\text{sym}}$, $V_{\text{asym}}$) and $P_{\text{RF}}$ (see inset of Figure 2b) thus excluding a possible contribution from the thermal effects. Other spurious galvanometric effects, such as the interplay between the stray RF currents and the oscillating magnet, related to either the anisotropic magnetoresistance (AMR) or to the planar or anomalous Hall effect (PHE or AHE), may come into play at this stage[19,30]. Each process is characterized by its own angular dependent signature. Due to the specific geometry of our coplanar wave guide (CPW), the RF excitation field $h_{RF}$ is maintained parallel to the sample stripe, with an angle ($\phi_0$) between the stripe direction and the in-plane DC field varying from 0 to 360° (see inset of Figure 2c). As mentioned in Ref.[19], the symmetric voltage contribution to the ISHE has the following angular dependence: $V_{sym}^{ISHE} = A_{ISHE} \frac{\omega_\phi}{\omega(\omega_\theta+\omega_\phi)^2} \cos^3 \phi_0$, where $\omega$, $\omega_\theta$ and $\omega_\phi$ are the angular frequency of the magnetization precession at resonance and its respective components in polar coordinates. By contrast, the AMR and AHE-like signals are characterized by the following shapes, $V_{sym}^{rect} = A_{rect} \cos \phi_0 \frac{\omega_\phi \Delta\rho_{AMR}\left(\cos 2\phi_0 \text{Im}[j_{rf}^{x*}]+\sin 2\phi_0 \text{Im}[j_{rf}^{y*}]\right)+\omega\rho_{AHE}\text{Re}[j_{rf}^{x*}]}{\omega(\omega_\theta+\omega_\phi)}$. $\text{Im}[j_{rf}^{x*}]$, $\text{Im}[j_{rf}^{y*}]$ and $\text{Re}[j_{rf}^{x*}]$ are respectively the imaginary and real parts of the complex conjugate of the RF current passing through the sample along the respective *x* and *y* directions. Without giving more details, Figure 2c and 2d display the results of the angular fit for two samples with different Si interlayer thicknesses (3μm and 37nm, respectively). The particular ISHE contribution related to a $\cos^3 \phi_0$ shape appears very clearly in both cases. For the 3μm thick Si sample, the $A_{ISHE}$ coefficient due to the ISHE contribution was found to be 3.28×10$^{-8}$V, while the $A_{rect}$ coefficient due to the AMR contribution is found to be much smaller around 9.6×10$^{-9}$V. Those values show that the spin-current generated by spin pumping can be electrically detected. Finally, the spin-current density $j_s^0$ detected in Pt is estimated to be $j_s^0$=9.56×10$^{-13}$J/m$^2$ giving a rough approximate value of the effective spin-mixing conductance of the order of $g_{eff}^{\uparrow\downarrow}$=1.6×10$^{17}$/m$^2$ for the 3μm thick Si sample (see Supplementary Note 3).

In addition, a control sample capped with a thin Cu layer (characterized by a spin Hall angle $\theta_{SHE}$ 30 times lower than Pt from Ref.[31]) instead of Pt, leads to a much smaller signal in the EMF symmetric component (see Supplementary Note 4), which also excludes any thermal effect to be responsible for the electrical signal. This further proves that the symmetric component of the EMF signal observed in the CoFeB/MgO/Si/Pt structure is not related to the ISHE from the Si interlayer but it is indeed strongly correlated to the ISHE from the top Pt layer.

**Si and MgO thickness dependence of damping and EMF signal**

In the following, systematic SP-ISHE experiments were performed to explore the spin-current injection in hybrid CoFeB/MgO/Si/Pt vertical structures. Both Si and MgO thicknesses were varied.

A) *Variation of the Si interlayer thickness with a fixed 2.2 nm thick MgO barrier*

We first consider a sample series in which the Si-thickness ($t_{Si}$) is varied (37nm, 124nm, 3µm and 10µm) while the same growth conditions are used for MgO (2.2nm), CoFeB (5.2nm) and Pt (6nm). Figure 3a displays the corresponding damping parameters in comparison with the reference samples. The damping parameter falls-off rapidly with increasing Si thickness, reaching a saturation of $4.5\times10^{-3}$ in the sample containing a 10µm thick Si layer. Figure 3b and 3c show the Si-thickness dependence of the $V_{sym}$ component in the EMF signal and the converted charge current $I_{sym}$, respectively. The EMF signals of samples containing different Si-thickness were measured at the same frequency with identical RF excitation power. As shown in Figure 3c, $I_{sym}$, normalized to the 37nm sample, falls off rapidly with increasing Si thickness highlighting an attenuation of the spin-signal for thicker Si layers. For the 10µm thick Si sample, the EMF signal is in the range of the noise thus proving that the origin of the signal comes from the spin-current after its propagation through the Si layer.

B) *Variation of the MgO thickness with a fixed 3µm thick Si interlayer*

A second series of samples was investigated in which the MgO thickness is varied between 1 and 3 nm while keeping the Si layer thickness fixed at 3µm. Figure 4b displays the variation of the damping with different MgO thicknesses ($t_{MgO}$=1, 2.2 and 3nm, respectively). The damping stays almost constant from 4.8 to $5.1\times10^{-3}$. However, surprisingly, the symmetric component of the acquired EMF voltage ($V_{sym}$) is pretty small at $t_{MgO}$=1nm and increases dramatically for thicker MgO samples ($t_{MgO}$=2.2nm and $t_{MgO}$=3nm), as shown in Figure 4a and 4b. This result is quite unusual since the spin-current is expected to decrease exponentially with the thickness of the tunnel barrier[32].

**Model of MgO thickness dependence of the charge accumulation at the MgO/Si interface**

Concerning this second set of experiments consisting of a fixed $t_{Si}$ and a varying $t_{MgO}$, the increase of the EMF signal could be related to a reduction of the Schottky barrier height (SBH) with increasing $t_{MgO}$. To clarify this issue, one needs to consider the charge transfer in a metal-insulator-semiconductor (MIS) structure involving the different charge reservoirs. If one introduces $\Delta = E_{CNL} - E_F$ as the energy difference between the Fermi level and the charge neutrality level (CNL) (Figure 4c), the SBH, $\Phi$ is given by $\Phi = \Phi_0 + \Delta$ where $\Phi_0$ is the nominal SBH when the Fermi level matches the CNL ($E_F = E_{CNL}$). Without metallic contact, $\Delta$ is non-zero (and negative) owing to the charge transfer between the conduction band (CB) and the interface states.

At equilibrium with the metal contact, $Q_{FM} = Q_{dep} - Q_{LS}$, where $Q_{FM}$ is the charge accumulated in the FM, $Q_{dep} = eN_D W$ is the total charge (hole) within the Si depletion layer with the width $W = \sqrt{\frac{2\varepsilon_{SC}(\Phi_0+\Delta)}{eN_D}}$ ($\varepsilon_{SC}$ is the dielectric permittivity of Si and $N_D$ is the doping concentration in Si), and $Q_{LS}$ is the charge (electron) accumulated at the MgO/Si interface (LS). $Q_{LS}$ may be expressed as $Q_{LS} = -eN_S\Delta$ assuming a flat band of density $N_S$. Due to the difference of work function between the CoFeB ($\Psi_{FM}$) and Si ($\Psi_{Si}$) with $\Delta_0 = \Psi_{FM} - \Psi_{Si}$, the total charge in the FM side can be expressed as $Q_{FM} = \frac{\varepsilon_B}{t_{MgO}}(\Delta_0 - \Delta)$, where $\varepsilon_B$ is the dielectric permittivity of MgO. This gives:

$$\frac{\varepsilon_B}{et_{MgO}}(\Delta_0 - \Delta) = N_D\sqrt{\frac{2\varepsilon_{SC}(\Phi_0+\Delta)}{eN_D}} + N_S\Delta \qquad (1)$$

and consequently the implicit relationship $\Delta$ vs. $t_{MgO}$ we are searching for. The following parameters have been used: $\varepsilon_B = \varepsilon_B^r \varepsilon_0 = 7.08 \times 10^{-11}$ F·m$^{-1}$, $\varepsilon_{SC} = \varepsilon_{SC}^r \varepsilon_0 = 1.06 \times 10^{-10}$ F·m$^{-1}$, $\Delta_0 \approx 0.1$ V (see Ref.[33]), $\Phi_0 = 0.25$ V (see Ref.[34]) and $N_D = 1 \times 10^{16}$ cm$^{-3}$. Note that in Ref.[35], the density of interface traps obtained by equivalent sputtered method was found to be close to 8×10$^{13}$cm$^{-2}$V$^{-1}$. Figure 4d displays the calculated results for $\Delta(t_{MgO})$ corresponding to different LS density $N_S$ in the range 1×10$^{14}$ - 1×10$^{16}$cm$^{-2}$V$^{-1}$. One observes that $\Delta$ decreases with $t_{MgO}$, thus leading to a decrease of the SBH with the MgO thickness. In addition, the higher density $N_S$ yields a smaller variation in $\Delta$ with $t_{MgO}$. This can be understood due to the pinning effect by LS at interface.

Our simulation shows a significant drop of SBH vs. $t_{MgO}$ by some tens of meV from its nominal value between 65-250meV[34,36]. In fact, the increase of $t_{MgO}$ leads to a decrease of the CoFeB/MgO/Si capacitance. This results in a reduction of the charge transfer between the two sides of the barrier and a decrease of the effective SBH. As a consequence, a larger spin-mixing conductance toward Pt ($g_{Tr}^{\uparrow\downarrow}$) via tunneling/hopping through the Schottky barrier and the depletion layer by thermal activation leads to an increase of the EMF signal. In addition, the small difference in damping between the 3μm and 10μm thick Si samples seems to indicate that most of the spin-flips or mixing occur on the LS at the MgO/Si interface ($g_{LS}^{\uparrow\downarrow}$) but not in Pt ($g_{Tr}^{\uparrow\downarrow}$) with the following relationship $g_{LS}^{\uparrow\downarrow} = \frac{e^2 N_s}{\tau_{sf}^{LS}} > g_{Tr}^{\uparrow\downarrow}$, where $\tau_{sf}^{LS}$ is spin lifetime in localized states. The finding of a constant value for $\Delta\alpha$ vs. $t_{MgO}$ emphasizes that the spin-flip process mainly occurs in an almost flat LS band.

**Evidence of spin pumping in localized states at MgO/Si interface**

One of our remarkable results is the determination of an effective spin-mixing conductance $g_{eff}^{\uparrow\downarrow}$ lying in the $1.6\times10^{17}$-$7.6\times10^{17}$/m² range for 3μm and 37nm thick Si samples, respectively. This is about two orders of magnitude smaller than values obtained for Co/Pt[22], Fe/Pt[37] or CoFeB/Pt[38] metallic interfaces ($g_{eff}^{\uparrow\downarrow}$~2-8×10¹⁹/m²). However, this remains in the same range as the one given for NiFe/GaN:Si junctions (1.38x10¹⁸/m²) extracted from the same spin-pumping techniques[39] and it appears larger than the one expected with a tunnel barrier[37]. On the other hand, a reference sample with the layer stack *n*-Si//MgO (2.2nm)/CoFeB (5.2nm)/MgO (2.2nm)/Pt (6nm), free of any Si interlayer, is used to compare with the sample containing a Si interlayer. From Figure 3a and 5a, one can find the following hierarchy $\alpha_{CoFeB/Pt} > \alpha_{CoFeB/MgO/Si/Pt} > \alpha_{CoFeB/MgO/Pt}$ together with $V_{CoFeB/Pt}^{sym} > V_{CoFeB/MgO/Si/Pt}^{sym} > V_{CoFeB/MgO/Pt}^{sym}$. It becomes obvious that the Si interlayer results in a promotion of a spin-current at the MgO/Si interface thus yielding a much larger EMF signal.

To explain the enhancement of the spin-current with the Si interlayer, we discuss below four possible mechanisms: (*i*) a spin-pumping within the ferromagnet with its own spin-motive force, followed by a tunneling of the spin-current through the MgO barrier; (*ii*) a spin-pumping within the ferromagnet, followed by the propagation of the spin-current via hopping processes through the localized states embedded in the barrier; (*iii*) a spin-pumping into the conduction band of Si (Bloch states) by indirect proximity exchange coupling with the ferromagnet; (*iv*) a spin-pumping into localized states (*e.g.* at the interface between the Si semiconductor and the oxide barrier) by proximity exchange coupling with the ferromagnet. For the first scenario (*i*), the spin-pump efficiency scales with the spin-mixing conductance (electronic transmission). As evidenced in the Fe/MgO/Pt system[37], the insertion of MgO barrier results in an exponential decrease of the spin-mixing conductance, which cannot explain our results obtained by varying MgO thickness. The second scenario (*ii*) has recently been invoked to explain spin-pumping and spin-transfer torque FMR (STT-FMR) experiments in LaAlO$_3$/SrTiO$_3$ oxide systems [27,40]. As discussed by Wang *et al.*[27], it may consist in the propagation of the spin via oxygen vacancies in MgO and thermal hopping excitations[41,42]. Nevertheless, one would also observe a significant spin-to-charge conversion in our CoFeB/MgO/Pt sample, which is obviously not the case. This thus confirms the major role played by the semiconductor itself. The third scenario (*iii*) of spin-pumping into the conduction band of Si (Bloch states) by indirect proximity coupling with the ferromagnet is in the same spirit as the occurrence of an oscillating exchange in Fe/MgO/Fe junctions[43]. However, it is not expected to be efficient in the present situation because of the reduced localization of the Bloch wavefunctions in the MgO barrier and in the Si depletion zone close to the interface. Indeed, the proximity exchange coupling field in a MgO-based magnetic tunnel junction is in the order of 10 mT for a 1-2 nm thick MgO barrier, which can create an exchange energy in the range of 1 μeV[44,45]. This appears to be too small to promote spin-pumping effect with a relatively thick MgO barrier (2-3nm) in our case. A larger weight of the density of the evanescent wavefunction inside the barrier would lead to a significant increase of the necessary coupling. The fourth scenario (*iv*) then considers the feasibility of such wavefunction density enhancement at the direct vicinity of barrier/SC interface. As schematically shown in Figure 5b, on the condition of a proximity exchange coupling larger than a threshold value of the order of $\hbar\omega$, a large spin-accumulation by SP may occur at the MgO/Si interface, established via the exchange proximity effects[46,47] or more probably through a chain of LS in the MgO barrier and the Si depletion layer[47]. The possibility of an enhancement of the indirect exchange via a resonant coupling between localized states and a FM contact has been already addressed for the issues of magnetic exchange[48,49] and spin-pumping effect between a FM insulator and a non-magnetic metal[50,51]. The scenario (*iv*) is also supported by the recent observations of the existence of an interlayer exchange coupling in relatively thick FM/MgO/FM junctions (1-3nm) revealed by a minimum value around 2nm[52]. This conclusion is reinforced by our results in the MgO thickness series samples showing up an increase of the ISHE signal with larger MgO barriers.

**Mechanism of spin pumping in localized states**

Our description for the spin pumping in LS (see also detailed model in Supplementary Note 5) is equivalent to the theoretical approaches derived recently by using Green's function techniques[46,53]. The spin-accumulation is driven by a spatially uniform dynamic exchange field. Using a perturbation treatment in a local rotating frame, the pumped spin-current is shown to exist down to coupling strength in the range of the RF-photon $\hbar\omega$ energy[46]. Here, the proof is based on the change of the spin-polarized carriers' Fermi energy on LS ($\Delta\epsilon_F^\sigma$) induced by the magnetization precession calculated to the third order by the perturbation technique, which can be expressed as:

$$\Delta\epsilon_F^\sigma = \mp \sin^2\theta \cdot \hbar\omega = \mp \left(\frac{h_{RF}}{H_{DC}}\right)^2 \hbar\omega \qquad (2)$$

where $\theta$ is the magnetization precession angle which is derived from the ratio between RF excitation field $h_{RF}$ and the DC magnetic field $H_{DC}$. The energy gain/loss is different and of the opposite sign between the two ↑ and ↓ spin states, leading to an out-of-equilibrium spin-accumulation in LS we are searching for. The spin-splitting or spin-accumulation can be obtained as $\Delta\mu_{sp} = \Delta\epsilon_F^\downarrow - \Delta\epsilon_F^\uparrow = 2\sin^2\theta \cdot \hbar\omega$ in the LS band. In the limit of a large spin-lifetime, the pumped spin-accumulation is independent of the exchange strength $J_{exc}$ once $J_{exc}$ is larger than $\hbar\omega$[46]. This leads to the standard expression of the spin-accumulation pumped in a metal[54]. The exact calculation of the transient spin dynamics on LS within the rotating exchange field gives $\Delta\mu_{sp} \propto \frac{\gamma_{exc}\left(\gamma_{exc}-\frac{\omega}{2}\right)}{(\gamma_{exc}-\omega)^2+\frac{1}{\tau_{sf}^2}}$, where $\gamma_{exc} = \frac{J_{exc}}{\hbar}$ is the (tunneling) exchange pulsation and $\tau_{sf}$ is the characteristic spin relaxation time. The latter expression indicates that the spin pumping works efficiently for a coupling $J_{exc} \geq \hbar\omega \approx 40\mu eV$, far below the typical on-site *s-d* coupling energy in ferromagnets (0.1 eV)[55]. One can estimate the maximum MgO barrier thickness meeting the requirement of a minimum exchange coupling ($40\mu eV$) for SP. The direct exchange $\Delta_{exc}$ is of the order of 1-2eV in transition metals. With the inserted MgO barrier, the exchange reduces approximately to $\Delta_{exc} \times T$, where $T$ is the transmission coefficient or transparency of the barrier. From Ref.[56] corresponding to CoFe/MgO/Si magnetic tunnel contact, one can find an increase of 4.5 order of magnitude in the CoFe/MgO/Si resistance when the MgO thickness increases from 0.5 to 2.5 nm. A reduction of the transmission by a factor 2.5x10$^4$ (reduction of the proximity exchange from 1 eV without tunnel barrier to 40 μeV in the case of a tunnel contact) then should correspond to a maximum of barrier thickness of 2.3nm. Owing to the existence of localized impurities or defects in the barrier (*e.g.* oxygen vacancies), the coupling strength can be even larger together with the maximum MgO barrier thickness.

The proposal of an enhanced spin-susceptibility and subsequent spin-transmission in the FMR regime has been recently theoretically addressed by Harmon *et al.*[57]. Our calculations developed in Supplementary Note 5 (Supplementary Figure 4) reveal that such enhancement in the spin response is linked to the large cone precession angle of the localized rotating spin in LS free of any damping (for a s-band) close to the resonance. Such enhancement in the spin susceptibility response means that the rotating spin angle deviates much more from the local field direction near the resonance, phenomena which is well known from the theory of FMR. The detailed equations giving the dynamics of SP in LS and given in the Supplementary Note 5 to clearly explain such trajectory near the resonance. The present model demonstrates the possibility to enhance the spin-currents by FMR spin-pumping methods in a complex structure or device made of magnetic tunnel contacts involving semiconductors.

One has to emphasize that the process of spin-injection by FMR spin-pumping differs from a standard two-point electrical spin-injection method in two major points. *i*) The spin-pumping can generate a spin-accumulation but not a spin-polarized current like by the electrical spin-injection method. Although it is sensitive to the spin-backflow process, the SP is less impacted by the so-called impedance mismatch problem[58,59] which limits the spin accumulation to the level of the spin-flip resistance of the FM. Unlike the electrical spin-injection method, the spin-accumulation generated by spin-pumping is only limited by the spin-mixing conductance (surface conductance). *ii*) Moreover, the spin-injection by SP may be favored via the presence of LS thanks to the pump action discussed above. On the contrary, those localized states may play the role of passive reservoir for spin-flip when they are involved in the electronic transport between a source and a drain[60], limiting thus the spin signal. The model of SP in LS presented here [scenario (*iv*)] can also be extended to explain the spin pumping in oxide LaAlO$_3$/SrTiO$_3$ two-dimensional electron gas[27,40] and FM/topological insulator interface with strong spin-orbit coupling effect[26]. Indeed, it can be shown that the same model with inclusion of spin-orbit terms may introduce a lateral charge current or galvanometric effects as observed in recent experiments[27,40,61].

**Estimation of the spin diffusion length in Si**

We now focus on the spin-current diffusion and relaxation process by considering three different spin-flip mechanisms: the spin-flips in the localized states at the MgO/Si interface, the spin-flips in Si leading to a finite spin-diffusion length (SDL) and the spin-flips in the top Pt layer. The particular variation of the damping enhancement involves two different contributions: $\Delta\alpha \propto g_{sf}^{\uparrow\downarrow} = g_{LS}^{\uparrow\downarrow}(t_{Si} = \infty) + g_{Tr}^{\uparrow\downarrow}(t_{Si})$. One is a constant term $g_{LS}^{\uparrow\downarrow}(t_{Si} = \infty)$ associated to the spin-flips in LS and the other term is a $t_{Si}$ dependent contribution $g_{Tr}^{\uparrow\downarrow}(t_{Si})$ describing the spin-escape towards the top Pt layer followed by the spin-relaxation through the Si interlayer. For the first set of data, varying $t_{Si}$ while keeping a fixed $t_{MgO}$, both damping parameter and ISHE signal decrease with $t_{Si}$. This is related to a drop of the spin-escape rate as well as the spin-mixing conductance term $g_{Tr}^{\uparrow\downarrow}(t_{Si})$ at larger $t_{Si}$. The increase of $t_{Si}$ results in a smaller number of spin-flips in Pt as well as a lower ISHE signal. Importantly, from those data, one can extract the characteristic spin diffusion length (SDL) in Si by taking into account the propagation of the spin-current within the ferromagnetic metal/semiconductor/heavy metal (FM/SC/HM) structures (see Supplementary Note 6). Two different methods may be applied for the extraction of the SDL.

*i)* The first method would consist in a fit of the variation of the damping parameter as well as the spin-mixing conductance $g_{Tr}^{\uparrow\downarrow}(t_{Si})$ term *vs.* $t_{Si}$ (Figure 3a) through its dependence on the SDL in Si. However, the exact shape of $g_{Tr}^{\uparrow\downarrow}(t_{Si})$ depends on the profile of the Schottky barrier height and its transmission *vs.* $t_{Si}$ which make the procedure not straightforward. Moreover, possible processes of spin pumping in chain in Si like that described in Ref.[53] make the searched dependence quite uncertain.

*ii)* The second method to extract SDL consists in the analysis of the $t_{Si}$ dependence of the ISHE signal according to the following formula: $\frac{I_{ISHE}}{g_{Tr}^{\uparrow\downarrow}\Delta\mu_{sp}} = \frac{r_{Si}^{S\infty}}{r_{Si}^{S\infty}\cosh\left(\frac{t_{Si}}{\lambda_{Si}}\right) + r_{Pt}^{S\infty}\coth(\frac{t_{Pt}}{\lambda_{Pt}})\sinh(\frac{t_{Si}}{\lambda_{Si}})}$, where $\Delta\mu_{sp}$ is the spin-splitting of the electrochemical potential in the LS, either injected from the FM metal or generated by SP at the MgO/Si interface. Figure 3d schematically shows the decay of $\Delta\mu_{sp}$ in the *n*-Si interlayer. $r_{Si}^{S\infty}$ and $r_{Pt}^{S\infty}$ are the respective spin resistance of Si and Pt when their thicknesses approach infinite. One obtains $I_{ISHE} \approx g_{Tr}^{\uparrow\downarrow} \frac{\Delta\mu_{sp}}{\cosh\left(\frac{t_{Si}}{\lambda_{Si}}\right)}$ since $r_{Si}^{S\infty} \gg r_{Pt}^{S\infty}$, which depends on the SDL $\lambda_{Si}$ we are searching for.

By using the second method (see details in Supplementary Note 7), we have finally extracted the SDL to be 2.0±0.3μm at room temperature, *i.e.* about two times larger than the values reported for the heavily *n*-doped Si (~2×10$^{19}$cm$^{-3}$) measured in a non-local configuration in the lateral devices (~1μm at RT)[11,12]. The larger SDL obtained in the vertical device could be due to two reasons: *i)* a reduced impurity scattering related to the lower doping concentration (1×10$^{16}$cm$^{-3}$) in our vertical device[62]; *ii)* the lack of surface or interface scattering in the vertical spin-current propagation channel compared to the lateral device structure[13]. The spin relaxation time $\tau_s$ may be estimated via the standard relationship $\lambda_{Si} = \sqrt{D\tau_s}$, where *D* is the spin diffusion constant. With $D$=6.5cm$^2$/s [9], one obtains $\tau_s$=6.2ns, a value much larger than the one obtained with a 3-terminal Hanle measurement in heavy *n*-doped Si (2×10$^{19}$cm$^{-3}$, 142ps at RT)[63] and three times larger than that obtained from the non-local measurements (~2ns at RT)[11,12]. Nevertheless, this value is in pretty good agreement with the theoretic calculation[64] using a realistic pseudopotential model (~8ns at RT) by taking into account the Elliott-Yafet spin-relaxation mechanism[65].

In conclusion, the relatively large spin-mixing conductance measured in our CoFeB/MgO/Si/Pt vertical device cannot be explained neither by a tunneling spin-current through the MgO barrier from the spin-accumulation in the FM contact by SP or by a spin-pumping in the Si conduction band states. The relevant scenario is the spin pumping in a very high density of LS at the MgO/Si interface via indirect exchange through the MgO barrier with the FM part. In that case, a small indirect exchange strength of the order of 40 μeV is sufficient to spin-pump a large amount of spin in the LS band, which then consequently generates a spin-current propagating into Si. Without the LS band, the spin-current generated by FM decays evanescently though the MgO barrier, as already evidenced in the Fe/MgO/Pt system[37]. In other words, an evanescent exchange coupling is much more efficient to generate the spin-current and spin-accumulation than an evanescent spin-current generated directly from the FM side.

**Conclusion**

In summary, by combining spin pumping and inverse spin Hall effect measurements, we have successfully demonstrated the efficient spin-current injection and transport at room temperature in CoFeB/MgO/Si/Pt vertical devices fabricated by ultra-high vacuum wafer-bonding technique. The spin-diffusion length in *n*-Si is determined to be 2.0±0.3μm at room temperature by studying a series of samples in which the Si interlayer thickness is varied. More importantly, our results evidence that the localized quantum electronic states at the MgO/Si interface play a major role to establish the necessary out-of-equilibrium spin-population which allows for efficient spin-current injection into Si by the spin-pumping process. The successful fabrication of metal-semiconductor-metal vertical structures represents an important breakthrough for semiconductor spintronics allowing investigations of the spin-current transport properties within a broad semiconductor thickness ranging from very thin (<40nm) to very thick (>300μm) layers.

**Methods**

*Sample preparation:*

In order to obtain CoFeB/MgO/Si/Pt vertical multi-layer structures, we employ the technique of ultra-high vacuum wafer-bonding. This technique has been already successfully used to study magnetic tunnel transistor[66] and hot electron spin-injection and transport in Si[16,17]. Firstly, MgO/$Co_{40}Fe_{40}B_{20}$ bilayer was sputtered on one *n*-Si wafer and one SOI wafer together in the sputtering system. The two wafers were then bonded together in the vacuum ($1\times10^{-8}$mbar) immediately after deposition. Chemical etching procedures were then used to remove the SOI Si handle layer and the $SiO_2$ box layer. Finally, a 6nm thick Pt layer was deposited on the top of the Si device layer. The final size of the structure is $1\times1cm^2$. The detailed fabrication procedure can be found in the Supplementary Note 1. To extract the important parameter of spin diffusion length, SOI wafers with different Si thicknesses were used. The doping in the Si device layer is about $1\times10^{16}cm^{-3}$, which gives a resistivity in the range of 1-10 Ω·cm. In all fabrication process, no UV or e-beam lithography was used to pattern the injector, channel and detector. The final structure of the layer stack is $n$-Si//MgO($t_{MgO}$)/CoFeB(5.2)/MgO($t_{MgO}$)/n-Si($t_{Si}$)/Pt(6) (with thickness in nanometer), as schematically shown in Figure 1a.

*FMR and ISHE measurements:*

FMR and EMF voltage measurements were performed at room temperature with samples set on CPW with Pt layer faced down. For FMR measurements, samples with a large size (10×5mm$^2$) were used. To eliminate the impact of the anisotropy of the magnetic layer, the measurement is averaged upon four sets by rotating the sample every 45°. The FMR measurement is performed in transmission mode via the field modulation and lock-in detection method. In general, the frequency range is kept between 4 and 18 GHz and the RF power is 20 dBm. For the spin pumping and EMF voltage measurement, the sample is cut into stripes of 10x1mm$^2$. To measure the EMF voltage, Au/Ti contacts were deposited by using a mask on the top Pt layer at the two ends of the sample stripe. As schematically shown in the inset of Figure 2c, the RF field generated by CPW is fixed to be parallel to the sample stripe. To measure the angular dependence of the ISHE, the ensemble stripe sample and CPW rotate together related to the external DC magnetic field direction and the angle ($\phi_0$) between the sample stripe direction and the DC magnetic field can vary from 0° to 360°. The EMF voltage measurement is performed with the frequency modulation and lock-in detection method.

**Acknowledgement:**

We thank Guillaume Sala and Alain Jacques for their help of developing UHV wafer-bonding technique. We also acknowledge the discussion with Prof. Mingwei Wu on spin transport in Si and the discussion with Prof. Mingcan Hu on spin pumping and ISHE. This work is supported by the joint French National Research Agency (ANR)-National Natural Science Foundation of China (NNSFC) ENSEMBLE project (Grants No. ANR-14-CE26-0028-01 and No. NNSFC 61411136001), Chinese-French International Key Program, National Natural Science Foundation of China (NSFC No.51620105004) and by the French PIA project "Lorraine Université d'Excellence" (Grant No. ANR-15-IDEX-04-LUE). C.C. acknowledges the supported by CNPq, National Council for Scientific and Technological Development-Brazil. A.D. acknowledges PhD funding from Region Lorraine. The experiments were performed using equipment from the platform TUBE–Davm funded by FEDER (EU), ANR, the Region Lorraine and Grand Nancy.

**Figures**

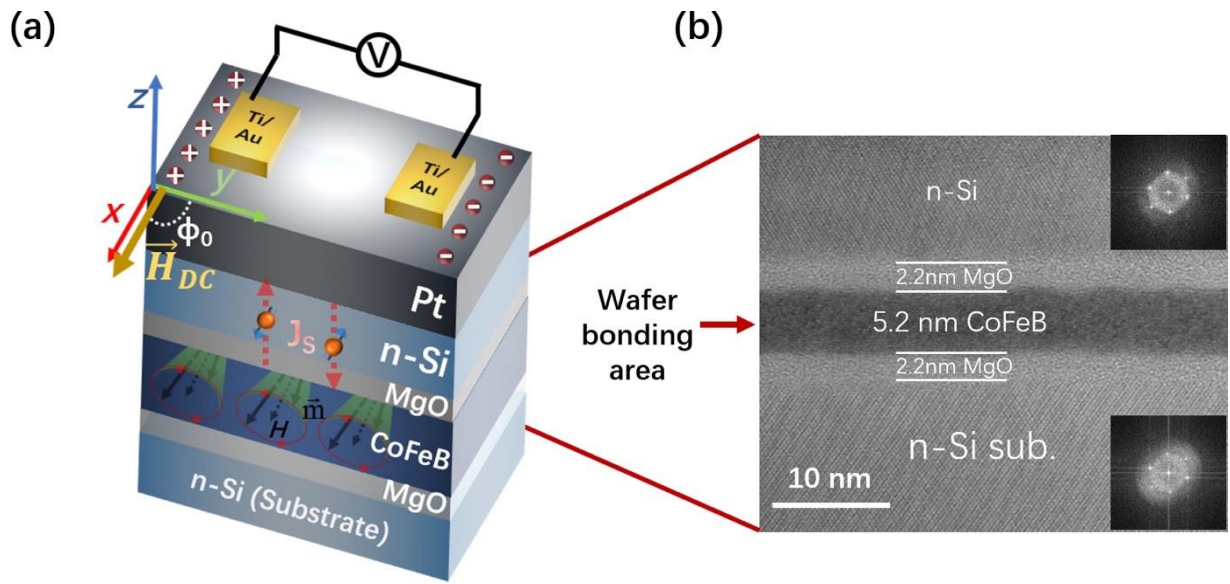

**Figure 1.** Device structure. (a) Schematic drawing of the vertical metal-Si-metal stack structure used for FMR and EMF voltage measurements. It also illustrates the process of spin current ($J_s$) injection/conversion by spin pumping of the CoFeB layer and ISHE of the top Pt layer, respectively. (b) HR-STEM image of the Si/MgO/CoFeB/MgO/Si wafer bonding region. Insets: top and bottom FFT diffraction patterns in the Si interlayer and the Si substrate, respectively.

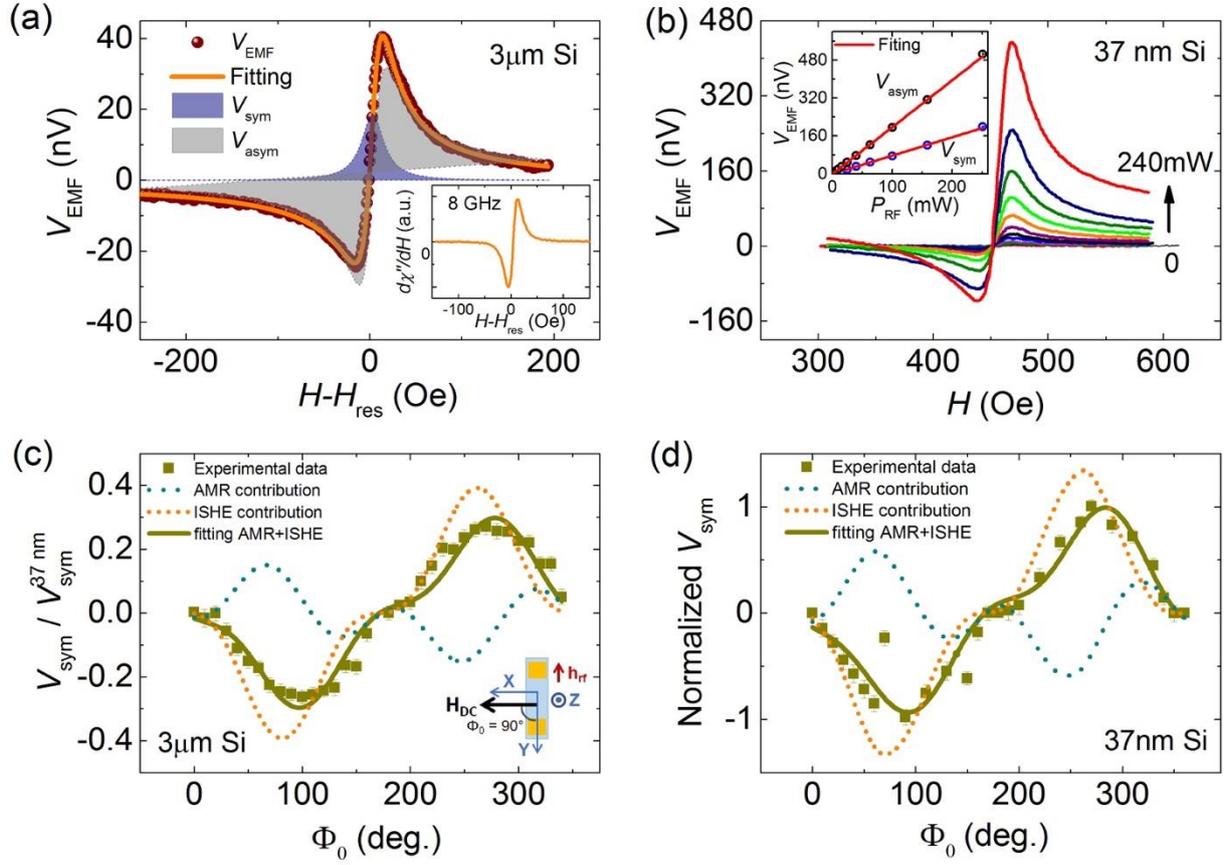

**Figure 2.** Evidence of spin-current generated by spin pumping. (a) EMF voltage measured with spin-pumping excitation in the CoFeB (5.2nm)/MgO (2.2nm)/Si (3μm)/Pt (6nm) sample. The symmetric and asymmetric components are separated with Lorentzian and anti-Lorentzian fitting of the spectrum, respectively. Inset: FMR spectra of the same sample with the same spin-pumping condition (8GHz). (b) EMF voltage with different excitation power in the sample CoFeB (5.2nm)/MgO (2.2nm)/Si (37nm)/Pt (6nm). Inset: The extracted $V_{sym}$ and $V_{asym}$ can be well linearly fitted. (c,d) Extracted $V_{sym}$ as a function of the angle $\phi_0$ between the sample and the DC magnetic field for (c): 3μm thick Si sample and (d): 37nm thick Si sample. The data was fitted by the model considering ISHE and AMR contributions. Inset of (c): Schematic configuration of angle dependence measurement.

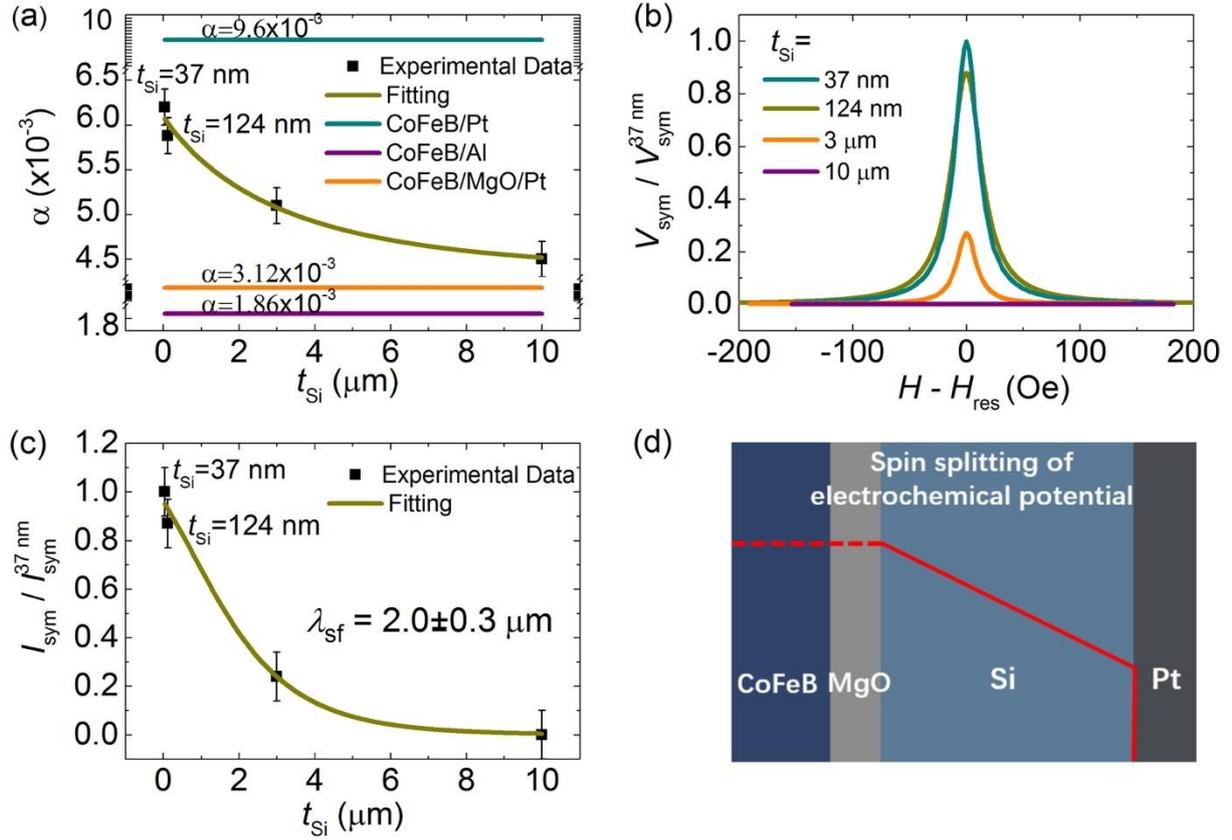

**Figure 3.** Si thickness dependent ISHE to extract the spin diffusion length. (a) Fitting of damping constant derived from FMR spectrum as a function of the *n*-Si interlayer thickness. The damping parameters of the reference samples of CoFeB/Pt, CoFeB/Al and CoFeB/MgO/Pt are also displayed. (b) Extracted ISHE symmetric components for the samples with different *n*-Si interlayer thicknesses of 37 nm, 124nm, 3μm and 10μm. The MgO layer thickness is fixed to be 2.2nm. The spin-pumping excitation frequency is 8GHz. (c) Fitting of symmetric component of ISHE current as a function of the *n*-Si interlayer thickness. (d) Schematics of the decay of the spin-splitting of the electrochemical potential ($\Delta\mu_{sp}$) in the *n*-Si interlayer.
20

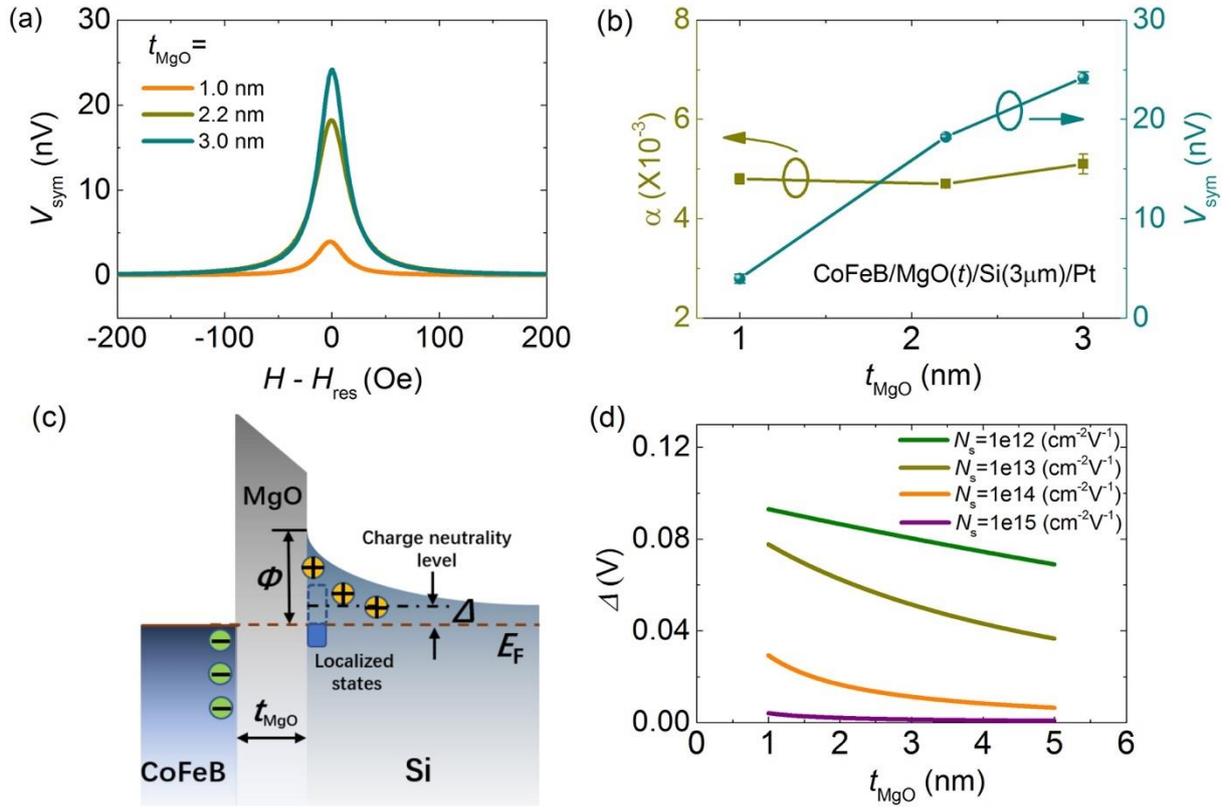

**Figure 4.** MgO thickness dependent ISHE and model of charge accumulation at MgO/Si interface. (a) Extracted EMF symmetric components for the samples with different MgO thicknesses of 1nm, 2.2nm and 3nm. The Si interlayer thickness is fixed to be 3μm. The spin-pumping excitation frequency is 8 GHz. (b) FMR damping and EMF symmetric component as a function of MgO thickness $t_{MgO}$ in the structure of CoFeB(5.2nm)/MgO($t_{MgO}$)/Si(3μm)/Pt(6nm). (c) Schematics to explain the decrease of charge transfer as well as Schottky barrier height (Φ) with thicker MgO barrier. (d) Simulation of the variation of the difference between charge neutrality level and Fermi level (Δ) as a function of MgO thickness with different LS density ($N_s$).

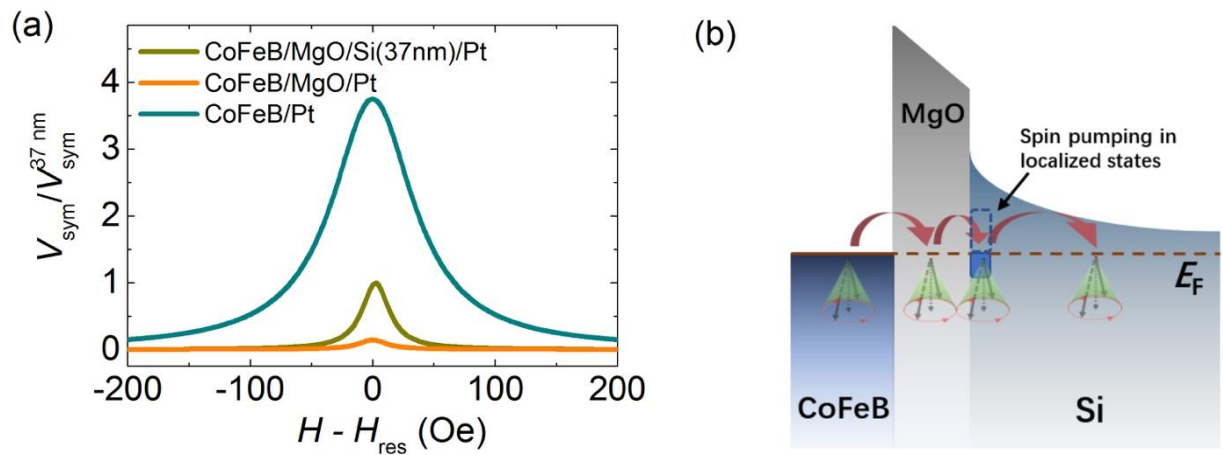

**Figure 5.** Evidence of spin pumping in the localized states at MgO/Si interface. (a) Comparison of EMF symmetric components for the samples with CoFeB/MgO/Si/Pt, CoFeB/MgO/Pt and CoFeB/Pt structures. CoFeB, Pt and MgO thickness were fixed to be 5.2nm, 6nm and 2.2nm, respectively. (b) Schematics of spin pumping via MgO/Si interface localized states.

**TOC figure**

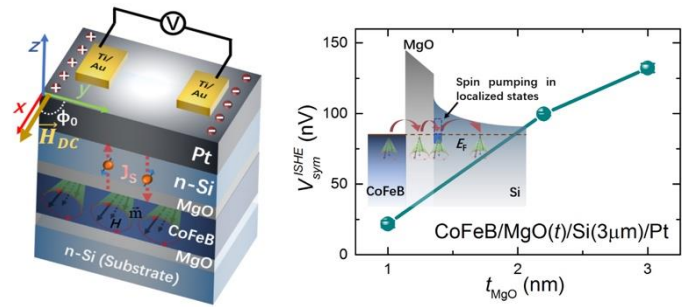

**TOC figure:** Schematic of spin-current injection in the metal-Si-metal vertical structure by spin pumping. We evidence pure spin-current generated by spin pumping in the localized states at the MgO/Si interface.

**References:**

1. Jansen, R. Silicon spintronics. *Nat. Mater.* **11.5**, 400 (2012).

2. Dyakonov, Michel I. & A. V. Khaetskii. *Spin Physics in Semiconductors* p211-243. (Springer, Berlin, 2008).

3. Tahan, C. & Joynt, R., Rashba spin-orbit coupling and spin relaxation in silicon quantum wells. *Phys. Rev. B* **71**, 075315 (2005).

4. Sigillito, A. J., Tyryshkin, A. M., Schenkel, T., Houck, A. A. & Lyon, S. A., All-electric control of donor nuclear spin qubits in silicon. *Nat. Nanotech.* **12**, 958–962 (2017).

5. Sasaki, T. et al. Local magnetoresistance in Fe/MgO/Si lateral spin valve at room temperature. *Appl. Phys. Lett.* **104**, 052404 (2014).

6. Van't Erve, O. M. J. et al. Electrical injection and detection of spin-polarized carriers in silicon in a lateral transport geometry. *Appl. Phys. Lett.* **91**, 212109 (2007).

7. Sasaki, T. et al. Temperature dependence of spin diffusion length in silicon by Hanle-type spin precession. *Appl. Phys. Lett.* **96**, 122101 (2010).

8. Le Breton, J. C., Sharma, S., Saito, H., Yuasa, S. & Jansen, R. Thermal spin current from a ferromagnet to silicon by Seebeck spin tunnelling. *Nature*, **475** (7354), 82. (2011).

9. Shikoh, E. et al. Spin-pump-induced spin transport in p-type Si at room temperature. *Phys. Rev. Lett.* **110**, 127201 (2013).

10. Ando, K. & Saitoh, E. Observation of the inverse spin Hall effect in silicon. *Nat. Commun.* **3**, 629 (2012).

11. Spiesser, A. et al. Giant spin accumulation in silicon nonlocal spin-transport devices. *Phys. Rev. Appl.* **8**, 064023 (2017).

12 Ishikawa, M. et al. Spin relaxation through lateral spin transport in heavily doped n-type silicon. *Phys. Rev. B* **95**, 115302 (2017).

13 Li, J. & Appelbaum, I. Modeling spin transport in electrostatically-gated lateral-channel silicon devices: Role of interfacial spin relaxation. *Phys. Rev. B* **84**, 165318 (2011).

14. Appelbaum, I., Huang, B. Q. & Douwe, J. M. Electronic measurement and control of spin transport in silicon. *Nature* **447**, 295 (2007).

15. Jang, H. J. et al. Non-ohmic spin transport in n-type doped silicon. *Phys. Rev. B* 78, 165329 (2008).

16. Lu, Y. & Appelbaum, I. Reverse Schottky-asymmetry spin current detectors. *Appl. Phys. Lett*. **97**, 162501 (2010).

17. Lu, Y., Li, J. & Appelbaum. I. Spin-polarized transient electron trapping in phosphorus-doped silicon. *Phys. Rev. Lett.* **106**, 217202 (2011).

18. Harder, M., Gui, Y. S. & Hu, C. -M. Electrical detection of magnetization dynamics via spin rectification effects. *Phys. Rep.* **661**, 1-59 (2016).

19. Iguchi, R. & Eiji, S. Measurement of spin pumping voltage separated from extrinsic microwave effects. *J. Phys. Soc. Jpn.* **86**, 011003 (2016).

20. Nakayama, H. et al. Geometry dependence on inverse spin Hall effect induced by spin pumping in $Ni_{81}Fe_{19}$/Pt films. *Phys. Rev. B* **85**, 144408 (2012).

21. Azevedo, A., Vilela-Leão, L. H., Rodríguez-Suárez, R. L., Santos, A. L. & Rezende, S. M. Spin pumping and anisotropic magnetoresistance voltages in magnetic bilayers Theory and experiment. *Phys. Rev. B* **83**, 144402 (2011).

22. Rojas-Sánchez, J-C. et al. Spin pumping and inverse spin Hall effect in platinum the essential role of spin-memory loss at metallic interfaces. *Phys. Rev. Lett.* **112**, 106602 (2014).

23. Dushenko, S. et al. Experimental demonstration of room-temperature spin transport in n-type germanium epilayers. *Phys. Rev. Lett.* **114**, 196602 (2015).

24. Watanabe, S. et al. Polaron spin current transport in organic semiconductors. *Nat. Phys.* **10**, 308 (2014).

25. Mellnik, A. R., et al. Spin-transfer torque generated by a topological insulator. *Nature* **511**, 449 (2014).

26. Jamali, M. et al. Giant Spin Pumping and Inverse Spin Hall Effect in the Presence of Surface and Bulk Spin-Orbit Coupling of Topological Insulator Bi2Se3. *Nano Lett.* **15**, 7126–7132 (2015).

27. Wang, Y. et al. Room-Temperature Giant Charge-to-Spin Conversion at the $SrTiO_3$−$LaAlO_3$ Oxide Interface. *Nano Lett.* **17**, 7659 (2017).

28. Soh, W. T., Peng, B., Chai, G. Z. & Ong, C. K. Note: Electrical detection and quantification of spin rectification effect enabled by shorted microstrip transmission line technique. *Rev. Sci. Instrum.* **85**, 026109 (2014).

29. Iguchi, R. & Saitoh, E. J. Measurement of Spin Pumping Voltage Separated from Extrinsic Microwave Effects. *Phys. Soc. Jpn.* **86**, 011003 (2017).

30. Bai, L. H. et al. Universal Method for Separating Spin Pumping from Spin Rectification Voltage of Ferromagnetic Resonance. *Phys. Rev. Lett.* **111**, 217602 (2013).

31. Wang, H. L. et al. Scaling of Spin Hall Angle in 3d, 4d, and 5d Metals from $Y_3Fe_5O_{12}$/Metal Spin Pumping. *Phys. Rev. Lett.* **112**, 197201 (2014).

32. Du, C. H. et al. Probing the Spin Pumping Mechanism Exchange Coupling with Exponential Decay in $Y_3Fe_5O_{12}$/Barrier/Pt Heterostructures. *Phys. Rev. Lett.* **111**, 247202 (2013)

33. Drummond, T. J. Work Functions of the Transition Metals and Metal Silicides. No. SAND99-0391J United States. Department of Energy, (1999).

34. Uhrmann, T. et al. Evaluation of Schottky and MgO-based tunnelling diodes with different ferromagnets for spin injection in n-Si. *J. Phys. D: Appl. Phys.* **42**, 145114 (2009).

35. Uhrmann, T. et al. Characterization of embedded MgO/ferromagnet contacts for spin injection in silicon. *J. Appl. Phys.* **103**, 063709 (2008).

36. Pu, Y. et al. Correlation of electrical spin injection and non-linear charge-transport in Fe/MgO/Si. *Appl. Phys. Lett.* **103**, 012402 (2013).

37. Mihalceanu, L. et al. Spin-pumping through a varying-thickness MgO interlayer in Fe/Pt system. *Appl. Phys. Lett.* **110**, 252406 (2017).

38. Conca, A. et al. Lack of correlation between the spin-mixing conductance and the inverse spin Hall effect generated voltages in CoFeB/Pt and CoFeB/Ta bilayers. *Phys. Rev. B* **95**, 174426 (2017).

39. Adhikari, R., Matzer, M., Martin-Luengo, A. T., Scharber, M. C. & Bonanni, A. Rashba semiconductor as spin Hall material: Experimental demonstration of spin pumping in wurtzite n-GaN:Si. *Phys. Rev. B* **94**, 085205 (2016).

40. Lesne, E. et al. Highly efficient and tunable spin-to-charge conversion through Rashba coupling at oxide interfaces, *Nat. Mater.* **15**, 1261–1266 (2016).

41. Lu, Y. et al. Spin-Polarized Inelastic Tunneling through Insulating Barriers. *Phys. Rev. Lett.* **102**, 176801 (2009).

42. Reyren, N. et al. Gate-Controlled Spin Injection at $LaAlO_3/SrTiO_3$ Interfaces. *Phys. Rev. Lett.* **108**, 186802 (2012).

43. Slonczewski, J. C. Conductance and exchange coupling of two ferromagnets separated by a tunneling barrier. *Phys. Rev. B* **39**, 6995 (1989).

44. Faure-Vincent, J. et al. Interlayer Magnetic Coupling Interactions of Two Ferromagnetic Layers by Spin Polarized Tunneling. *Phys. Rev. Lett.* **89**, 107206 (2002).

45. Nistor, L. E. et al. Oscillatory interlayer exchange coupling in MgO tunnel junctions with perpendicular magnetic anisotropy. *Phys. Rev. B* **81**, 220407(R) (2010).

46. Tatara, G. Green's function representation of spin pumping effect. *Phys. Rev. B* **94**, 224412 (2016).

47. Ou, Y. S. et al. Exchange-Driven Spin Relaxation in Ferromagnet-Oxide-Semiconductor Heterostructures. *Phys. Rev. Lett.* **116**, 107201 (2016).

48. Zhuravlev, M. Y., Tsymbal, E. Y. & Vedyayev, A. V. Impurity-Assisted Interlayer Exchange Coupling across a Tunnel Barrier. *Phys. Rev. Lett.* **94**, 026806 (2005).

49. Yang, H. X., Chshiev, M., Kalitsov, A., Schuhl, A. & Butler, W. H. Effect of structural relaxation and oxidation conditions on interlayer exchange coupling in Fe/MgO/Fe tunnel junctions, *Appl. Phys. Lett.* **96**, 262509 (2010).

50. Kajiwara, Y. et al. Transmission of electrical signals by spin-wave interconversion in a magnetic insulator. *Nature* **464**, 262 (2010).

51. Rezende, S. M., Rodriguez-Suarez, R. L. & Azevedo, A. Magnetic relaxation due to spin pumping in thick ferromagnetic films in contact with normal metals. *Phys. Rev. B* **88**, 014404 (2013).

52. Baker, A. A. et al. Spin pumping in magnetic trilayer structures with an MgO barrier. *Sci. Rep.* **6**, 35582 (2016).

53. Ahmadi, A. & Mucciolo, E. R. Microscopic formulation of dynamical spin injection in ferromagnetic-nonmagnetic heterostructures, *Phys. Rev. B* **96**, 035420 (2017).

54. Tserkovnyak, Y., Brataas, A., Bauer, G. E. W. & Halperin, B. I. Nonlocal magnetization dynamics in ferromagnetic heterostructures. *Rev. Mod. Phys.* **77**, 1375 (2005).

55. Kübler, J. *Theory of Itinerant Electron Magnetism.* (Oxford University Press, UK, 2009).

56. Saito, Y. et al. Correlation between amplitude of spin accumulation signals investigated by Hanle effect measurement and effective junction barrier height in CoFe/MgO/n+-Si junctions. *J. Appl. Phys.* **117**, 17C707 (2015).

57. Harmon, N. J. & Flatté, M. E. Theory of spin-coherent electrical transport through a defect spin state in a metal/insulator/ferromagnet tunnel junction undergoing ferromagnetic resonance. *Phys. Rev. B* **98**, 035412 (2018).

58. Fert, A. & Jaffrès, H, Conditions for efficient spin injection from a ferromagnetic metal into a semiconductor. *Phys. Rev. B* **64**, 184420 (2001).

59. Ando, K. & Saitoh, E. Observation of the inverse spin Hall effect in silicon. *Nat. Commun.* 3, 629 (2012).

60. Tran, M. et al. Enhancement of the Spin Accumulation at the Interface between a Spin-Polarized Tunnel Junction and a Semiconductor. *Phys. Rev. Lett.* 102, 036601 (2009).

61. Oyarzun, S. et al. Evidence for spin-to-charge conversion by Rashba coupling in metallic states at the Fe/Ge(111) interface, *Nat. Commun.* **7**, 13857 (2016).

62. Jansen, R. Silicon spintronics. *Nat. Mat.* **11**, 400 (2012).

63. Saroj, P. et al. Electrical creation of spin polarization in silicon at room temperature. *Nature* **462**, 491–494 (2009).

64. Cheng, J. L., Wu, M. W. & Fabian, J. Theory of the Spin Relaxation of Conduction Electrons in Silicon, *Phys. Rev. Lett.* **104**, 016601 (2010).

65. Yafet, Y. In *Solid State Physics*, edited by F. Seitz and D. Turnbull (Academic, New York, 1963), Vol. 14, p. 1.

66. Lu, Y. et al. Electrical control of interfacial trapping for magnetic tunnel transistor on silicon. *Appl. Phys. Lett.* **104**, 042408 (2014).